\documentclass[aps,pra,reprint,amsmath,amssymb,showpacs,floatfix]{revtex4-1}

\usepackage{graphicx,url}

\newcommand{\ket}[1]{\ensuremath{\vert{#1\rangle}}} 
\newcommand{\bra}[1]{\ensuremath{{\langle #1}\vert}}

\newcommand{\op}[1]{\hat{#1}}
\newcommand{\D}{\text{d}}
\newcommand{\I}{\text{i}}
\newcommand{\E}{\text{e}}
\providecommand{\abs}[1]{\left\lvert#1\right\rvert}
\newcommand{\bvec}[1]{\ensuremath{\mathbf{#1}}}

\usepackage[colorlinks,citecolor=blue,urlcolor=blue,linkcolor=blue,hyperindex,breaklinks]{hyperref}

\begin{document}

\title{Measuring the quantum state of a single system with minimum state disturbance}
\author{Maximilian Schlosshauer}
\affiliation{Department of Physics, University of Portland, 5000 North Willamette Boulevard, Portland, Oregon 97203, USA}

\begin{abstract} 
Conventionally, unknown quantum states are characterized using quantum-state tomography based on strong or weak measurements carried out on an ensemble of identically prepared systems. By contrast, the use of protective measurements offers the possibility of determining quantum states from a series of weak, long measurements performed on a single system. Because the fidelity of a protectively measured quantum state is determined by the amount of state disturbance incurred during each protective measurement, it is crucial that the initial quantum state of the system is disturbed as little as possible. Here we show how to systematically minimize the state disturbance in the course of a protective measurement, thus enabling the maximization of the fidelity of the quantum-state measurement. Our approach is based on a careful tuning of the time dependence of the measurement interaction and is shown to be dramatically more effective in reducing the state disturbance than the previously considered strategy of weakening the measurement strength and increasing the measurement time. We describe a method for designing the measurement interaction such that the state disturbance exhibits polynomial decay to arbitrary order in the inverse measurement time $1/T$. We also show how one can achieve even faster, subexponential decay, and we find that it represents the smallest possible state disturbance in a protective measurement. In this way, our results show how to optimally measure the state of a single quantum system using protective measurements. \\[-.1cm]

\noindent Journal reference: \emph{Phys.\ Rev.\ A\ }\textbf{93}, 012115 (2016), DOI: \href{http://dx.doi.org/10.1103/PhysRevA.93.012115}{10.1103/PhysRevA.93.012115}
\end{abstract}


\maketitle

\section{Introduction}

The characterization of unknown quantum states is an important experimental task and of great significance to quantum information processing \cite{Vogel:1989:uu,Dunn:1995:oo,Smithey:1993:lm,Breitenbach:az:1997,White:1999:az,James:2001:uu,Haffner:2005:sc,Leibfried:2005:yy,Altepeter:2005:ll,Lvovsky:2009:zz}. In conventional quantum-state tomography \cite{Vogel:1989:uu,Paris:2004:uu,Altepeter:2005:ll}, the quantum state is reconstructed from expectation values obtained from strong measurements of different observables, performed on an ensemble of identically prepared systems. An alternative approach to quantum-state measurement \cite{Lundeen:2011:ii,Lundeen:2012:rr,Fischbach:2012:za,Bamber:2014:ee,Dressel:2011:au} uses a combination of weak and strong measurements on an ensemble of identically prepared systems, together with the concept of weak values \cite{Aharonov:1988:mz,Duck:1989:uu}. However, since both approaches require an ensemble of identically prepared systems, they can only be said to reconstruct the quantum state in the statistical sense of measurement averages over an ensemble of systems presumed to have been prepared in the same quantum state. This raises the question of whether it might be possible to determine the quantum state of an individual system from measurements carried out not on an ensemble but on this single system only. Such single-system state determination would not only offer a conceptually transparent and rigorous version of quantum-state measurement, but also avoid time-consuming postprocessing and error propagation associated with quantum-state tomography \cite{Altepeter:2005:ll,Maccone:2014:uu,Dressel:2011:au}.

As long as one demands perfect fidelity of the state reconstruction and possesses no prior knowledge of the initial quantum-state subspace, then it is well known that single-system state determination is impossible \cite{Wootters:1982:ww,Ariano:1996:om}. However, if one weakens these conditions, then it has been shown that one can, in principle, measure the quantum state of a single system by using the protective-measurement protocol \cite{Aharonov:1993:qa,Aharonov:1993:jm,Aharonov:1996:fp,Alter:1996:oo,Dass:1999:az,Vaidman:2009:po,Gao:2014:cu}. Protective measurement allows for a set of expectation values to be obtained from weak measurements performed on the same single system, provided the system is initially in an (potentially unknown) nondegenerate eigenstate of its (potentially unknown) Hamiltonian. A defining feature of a protective measurement is that the disturbance of the system's quantum state during the measurement can be made arbitrarily small by weakening the measurement interaction and increasing the measurement time \cite{Aharonov:1993:jm,Dass:1999:az,Vaidman:2009:po}. Thus, a series of expectation values can be measured on the same system while the system remains in its initial state with probability arbitrarily close to unity. In this sense, one can measure the quantum state of a single system with a fidelity arbitrarily close to unity \cite{Aharonov:1993:qa,Aharonov:1993:jm,Aharonov:1996:fp,Dass:1999:az,Vaidman:2009:po,Auletta:2014:yy,Diosi:2014:yy,Aharonov:2014:yy}, providing an important complementary approach to conventional quantum-state tomography based on ensembles. Recently, the possibility of using protective measurement for quantum-state determination has attracted renewed interest \cite{Gao:2014:cu}, and protective measurement has been shown to have many related applications, such as the determination of stationary states \cite{Diosi:2014:yy}, investigation of particle trajectories \cite{Aharonov:1996:ii,Aharonov:1999:uu}, translation of ergodicity into the quantum realm \cite{Aharonov:2014:yy}, studies of fundamental issues of quantum measurement \cite{Aharonov:1993:qa,Aharonov:1993:jm,Aharonov:1996:fp,Alter:1997:oo,Dass:1999:az,Gao:2014:cu}, and the complete description of two-state thermal ensembles \cite{Aharonov:2014:yy}. 

The fact that each protective measurement has a nonzero probability of disturbing the quantum state of the measured system leads to error propagation and reduced fidelity over the course of the multiple measurements required to determine the set of expectation values \cite{Aharonov:1993:qa,Alter:1996:oo,Dass:1999:az,Schlosshauer:2014:tp,Schlosshauer:2014:pm,Schlosshauer:2015:pm}. Therefore, a chief goal when using protective measurement to characterize quantum states of single systems is the minimization of the state disturbance. However, the conventional approach of making the measurement interaction arbitrarily weak while allowing it to last for an arbitrarily long time \cite{Aharonov:1993:jm,Dass:1999:az,Vaidman:2009:po} is not only unlikely to be practical in an experimental setting but is also, as we will show in this paper, comparably ineffective. Here we will describe a dramatically more effective approach that allows one to minimize the state disturbance while keeping the strength and duration of the measurement interaction constant. In this way, we demonstrate how to optimally implement the measurement of an unknown quantum state of a single system using protective measurement. 

Our approach consists of a systematic tuning of the time dependence of the measurement interaction, such that the state disturbance becomes dramatically reduced even for modestly weak and relatively short interactions. While early expositions of protective measurement \cite{Aharonov:1993:qa,Aharonov:1993:jm,Aharonov:1996:fp} had hinted at the role of the time dependence of the measurement interaction, this role had not been explicitly explored and was instead relegated to a reference to the quantum adiabatic theorem \cite{Born:1928:yf}, which, as we will see in this paper, provides a condition that is neither necessary nor sufficient for minimizing the state disturbance in a protective measurement. Issues of time dependence of the protective-measurement interaction were first considered explicitly in Ref.~\cite{Dass:1999:az}, which estimated the effect of the turn-on and turnoff of the measurement interaction on the adiabaticity of the interaction. Recently, the case of finite measurement times in a protective measurement and its influence on the reliability of the measurement were studied \cite{Auletta:2014:yy,Schlosshauer:2014:tp}, and a framework for the perturbative treatment of time-dependent measurement interactions in a protective measurement has been developed and applied to specific examples \cite{Schlosshauer:2014:pm,Schlosshauer:2015:pm}. None of these existing studies, however, have shown how to systematically minimize the state disturbance in a protective measurement for the physically and experimentally relevant case of finite measurement times and interaction strengths, such that the reliability of the protective measurement can be maximized. 

Here we present a rigorous and comprehensive solution to this problem. Our results demonstrate how one can optimally measure the quantum state of an individual quantum system using protective measurements. In any future experimental implementation of protective quantum-state measurement, this will enable one to optimize the measurement interaction to produce a high fidelity of the quantum-state measurement. While our analysis is motivated by the goal of optimizing protective measurements, it also provides insights into the issue of state disturbance in any quantum measurement.
 
This paper is organized as follows. After a brief review of the basics of protective measurements involving time-dependent measurement interactions (Sec.~\ref{sec:prot-meas}), we first use a Fourier-like series approach to construct measurement interactions that achieve a state disturbance that decreases as $1/T^N$, where $T$ is the measurement time and $N$ can be made arbitrarily large by modifying the functional form of the time dependence of the measurement interaction using a systematic procedure (Sec.~\ref{sec:seri-appr-minim}). We also make precise the relationship between the smoothness of the measurement interaction and the dependence of the state disturbance on $T$. We then show that the measurement interaction can be further optimized, leading to an even faster, subexponential decay of the state disturbance with $T$, and we show that this constitutes the optimal choice (Sec.~\ref{sec:minim-state-dist}). These results are established by calculating the state disturbance from the perturbative transition amplitude to first order in the interaction strength. To justify this approach, we prove that this amplitude accurately represents the exact transition amplitude to leading order in $1/T$ (Sec.~\ref{sec:suff-first-order}).

\section{\label{sec:prot-meas}Protective measurement} 

We begin by briefly reviewing protective measurements and their treatment with time-dependent perturbation theory. In a protective measurement \cite{Aharonov:1993:qa,Aharonov:1993:jm,Aharonov:1996:fp,Dass:1999:az,Vaidman:2009:po,Gao:2014:cu}, the interaction between system $S$ and apparatus $A$ is treated quantum mechanically and described by the interaction Hamiltonian $\op{H}_\text{int}(t) = g(t)\op{O} \otimes \op{P}$, where $\op{O}$ is an arbitrary observable of $S$, $\op{P}$ generates the shift of the pointer of $A$, and the coupling function $g(t)$ describes the time dependence of the interaction strength during the measurement interval $0 \le t \le T$, with $g(t)=0$ for $t <0$ and $t >T$. The function $g(t)$ is normalized, $\int_{0}^{T} \D t\, g(t) =1$, which introduces an inverse relationship between the duration $T$ and the average strength of the interaction, so that the pointer shift depends neither on these two parameters nor on the functional form of $g(t)$. The spectrum $\{ E_n \}$ of $\op{H}_S$ is assumed to be nondegenerate and $S$ is assumed to be in an eigenstate $\ket{n}$ of $\op{H}_S$ at $t=0$. One can then show \cite{Aharonov:1993:qa,Aharonov:1993:jm,Dass:1999:az,Vaidman:2009:po,Gao:2014:cu} that for $T \rightarrow \infty$ the system remains in the state $\ket{n}$, while the apparatus pointer shifts by an amount proportional to $\bra{n}\op{O}\ket{n}$, thus providing partial information about $\ket{n}$. However, in the realistic case of finite $T$ and a corresponding non-infinitesimal average interaction strength, the system becomes entangled with the apparatus, disturbing the initial state \cite{Auletta:2014:yy,Schlosshauer:2014:tp,Schlosshauer:2014:pm,Schlosshauer:2015:pm}.  

To quantify this state disturbance, we calculate the probability amplitude $A_m(T)$ for finding the system in an orthogonal state $\ket{m}\not=\ket{n}$ at the conclusion of the measurement. We write $A_m(T)$ as a perturbative series, $A_m(T) = \sum_{\ell=1}^\infty A_m^{(\ell)}(T)$. Here $A_m^{(1)}(T)$ is the transition amplitude to first order in the interaction strength and the amplitudes $A_m^{(\ell)}(T)$ for $\ell \ge 2$ are the $\ell$th-order corrections to $A_m^{(1)}(T)$, where \cite{Sakurai:1994:om,Schlosshauer:2014:pm} 
\begin{align}\label{eq:g8fbvsv1}
A^{(\ell)}_{m}(T) &= \left(-\frac{\I}{\hbar}\right)^\ell \sum_{k_1,\hdots,k_{\ell-1}} O_{mk_1}O_{k_1k_2}\cdots O_{k_{\ell-1}n}\notag \\ & \quad \times\int_{0}^{T}  \D t' \, \E^{\I \omega_{mk_1} t'} g(t') \cdots \notag \\ & \quad \times\int_{0}^{t^{(\ell-1)}} \D t^{(\ell)} \,\E^{\I \omega_{k_{\ell-1} n} t^{(\ell)}} g(t^{(\ell)}).
\end{align}
Here $O_{ij}\equiv \bra{k_i} \op{O} \ket{k_j}$, and $\omega_{mn}\equiv (E_m-E_n)/\hbar$ is the frequency of the transition $\ket{n} \rightarrow \ket{m}$ \footnote{The full expression for $A^{(\ell)}_{m}(T)$ also contains contributions from the apparatus subspace \cite{Schlosshauer:2014:pm}. They are irrelevant to our present analysis.}. Of particular interest is the first-order transition amplitude $A_m^{(1)}(T)$, 
\begin{align}\label{eq:8aadhj7gr7ss82}
A_m^{(1)}(T) &= -\frac{\I}{\hbar} O_{mn} \int_{0}^{T} \D t\, \E^{\I \omega_{mn} t} g(t).
\end{align}
The total state disturbance is measured by the probability $\abs{\sum_{m\not=n} A_m(T)}^2$ of a transition to the subspace orthogonal to the initial state $\ket{n}$. Our goal is now to determine coupling functions $g(t)$ that minimize this transition probability.

\section{\label{sec:seri-appr-minim}Series approach to minimization of state disturbance}

Our first approach will consist of building up coupling functions $g(t)$ from sinusoidal components such that the coupling functions become increasingly smooth (in a sense to be defined below). We take $g(t)$ to be symmetric about $t=T/2$ and expand it in terms of the functions
\begin{equation}\label{eq:bvdhkjb678678}
f_n(t)= (-1)^{n+1}\cos\left[\frac{2n\pi (t-T/2)}{T}\right], \quad n=1,2,3,\hdots,
\end{equation}
which form an orthogonal basis over the interval $[0,T]$ for functions symmetric about $t=T/2$. That is, we write $g(t)$ as
\begin{equation}\label{eq:bvdhkjbvd}
g(t) = \begin{cases} \frac{1}{T} \left( 1 + \sum_{n=1}^N a_n f_n(t)\right), & 0 \le t \le T, \\ 0, & \text{otherwise}, \end{cases}
\end{equation}
where the coefficients $a_n$ are dimensionless and do not depend on $T$. Since $\int_0^T \D t \, f_n(t)=0$, the area under $g(t)$ is normalized as required. The dominant contribution comes from the $f_1(t)$ term describing a gradual increase and decrease. The terms $f_n(t)$ for $n \ge 2$ represent sinusoidal components with multiple peaks that we will now use to suitably shape the basic pulse represented by $f_1(t)$.

We will first consider the first-order transition amplitude $A_m^{(1)}(T)$ given by Eq.~\eqref{eq:8aadhj7gr7ss82}, and then subsequently justify this approach by showing that higher-order corrections $A_m^{(\ell\ge 2)}(T)$ do not modify the results. Equation~\eqref{eq:8aadhj7gr7ss82} shows that the coupling-dependent part of $A_m^{(1)}(T)$ is represented by the Fourier transform $G(\omega T) = \int_0^T \D t\, \E^{\I \omega t} g(t)$ of $g(t)$, where $\omega\equiv \omega_{mn}$. Thus, to quantify the state disturbance we evaluate the Fourier transform of $g(t)$ given by Eq.~\eqref{eq:bvdhkjbvd}, 
\begin{align}\label{eq:bvdhkjbvd0}
G(\omega T) = \frac{2\E^{\I \omega T/2}}{\omega T} \sin\left( \omega T/2 \right) \left[ 1 - \sum_{n=1}^N \frac{a_n}{1-(2\pi n/\omega T)^2 }\right],
\end{align}
where $\omega T$ is a dimensionless quantity that measures the ratio of the measurement time to the internal timescale $\omega^{-1}$ associated with the transition $\ket{n}\rightarrow\ket{m}$. In physical situations, $\omega^{-1}$ typically represents atomic timescales and we may safely assume that $\omega T \gg N$. Then we can write Eq.~\eqref{eq:bvdhkjbvd0} as a power series in $1/\omega T$,
\begin{equation}\label{eq:uuuun}
G(\omega T) =   \frac{2\E^{\I \omega T/2}}{\omega T} \sin\left( \omega T/2 \right)\left[ 1  - \sum_{k=0}^\infty \sum_{n=1}^N a_n \left(\frac{2\pi n}{\omega T}\right)^{2k} \right].
\end{equation}

To minimize the state disturbance, we want $G(\omega T)$ to decay quickly with $T$ from its initial value of 1 at $T=0$. For the constant-coupling function $g(t)=1/T$ (all $a_n=0$), which describes a sudden turn-on and turnoff, we obtain $A_m^{(1)}(T) \propto 1/\omega T$, where the $T$ dependence is due to the fact that the average interaction strength is proportional to $1/T$. Clearly, we must have $\omega T \gg 1$ to achieve small state disturbance. For arbitrary coefficients $a_n$, $A_m^{(1)}(T)$ is still of first order in $1/\omega T$. Equation~\eqref{eq:uuuun} shows that we may increase the order of the leading term in $1/\omega T$ by imposing the conditions 
\begin{align}\label{eq:conds}
\sum_{n=1}^N a_n=1, \quad \sum_{n=1}^N a_n n^{2k}=0, \quad 1 \le k \le N-1,
\end{align}
which define a set of $N$ linearly independent coupled equations for $N$ coefficients $a_n$ with a unique solution $\bvec{a}_N=(a_1,\hdots,a_N)$; e.g., $\bvec{a}_1=\left(1 \right)$, $\bvec{a}_2=\left(\frac{4}{3},-\frac{1}{3}\right)$, $\bvec{a}_3=\left(\frac{3}{2},-\frac{3}{5},\frac{1}{10}\right)$, etc. Using the solution $\bvec{a}_N$, $A_m^{(1)}(T)$  to leading order in $1/\omega T$ becomes [see Eqs.~\eqref{eq:8aadhj7gr7ss82}  and \eqref{eq:uuuun}]
\begin{align}\label{eq:uuuun2233}
\widetilde{A}_m^{(1)}(T)&=   -\frac{2\I}{\hbar} O_{mn} \E^{\I \omega T/2} \sin\left( \omega T/2 \right)\left(2\pi\right)^{2N} \notag \\ & \quad \times \left(\sum_{n=1}^N a_n n^{2N}\right) \left(\frac{1}{\omega T}\right)^{2N+1},
\end{align}
where the tilde indicates leading-order expressions. This amplitude is of order $(\omega T)^{-(2N+1)}$. 

\begin{figure}
\includegraphics[scale=.85]{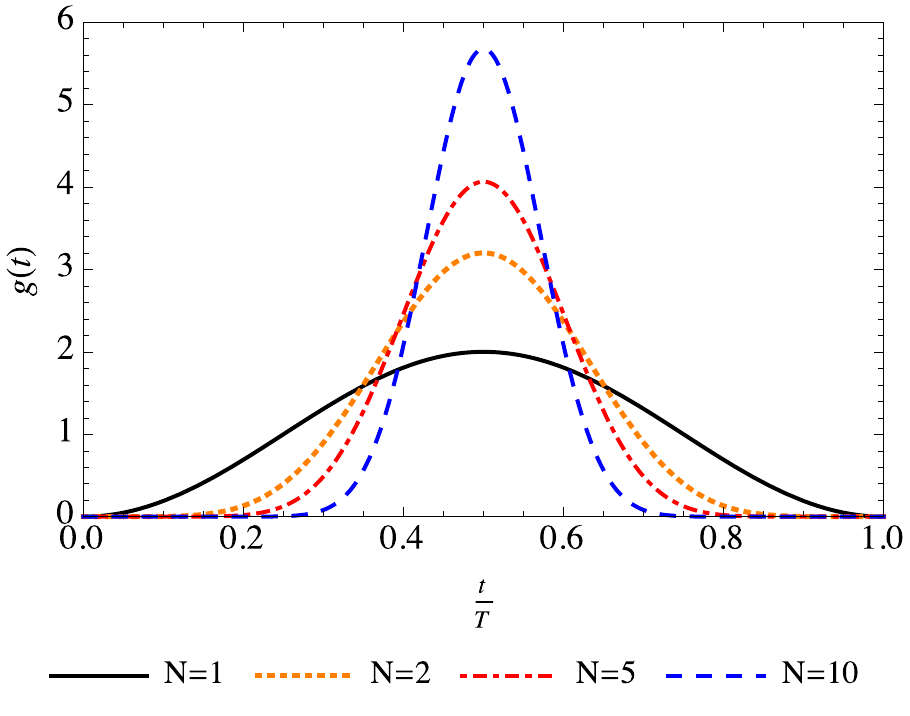}
\caption{\label{fig:g}(Color online) Coupling functions determined from the conditions in Eq.~\eqref{eq:conds} for different numbers $N$ of sinusoidal components [Eq.~\eqref{eq:bvdhkjb678678}]. The horizontal axis is in units of the measurement time $T$ and the vertical axis is in units of $1/T$.}
\end{figure}

\begin{figure}
\begin{flushleft}
{\small (a)}
\end{flushleft}

\vspace{-.3cm}

\includegraphics[scale=.9]{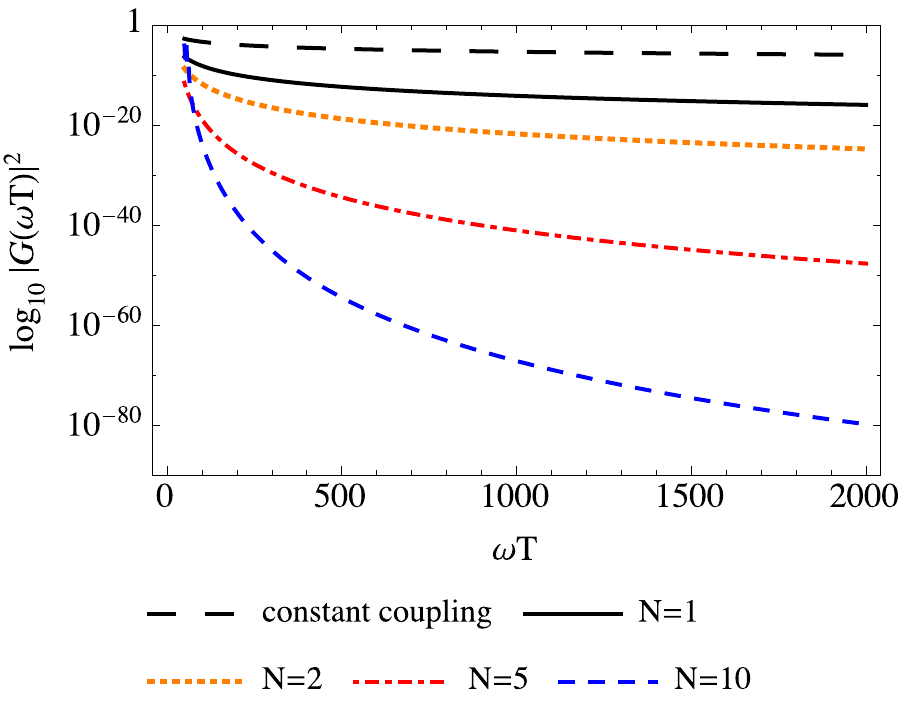}

\vspace{-.7cm}

\begin{flushleft}
{\small (b) }
\end{flushleft}

\vspace{-.3cm}

\includegraphics[scale=0.85]{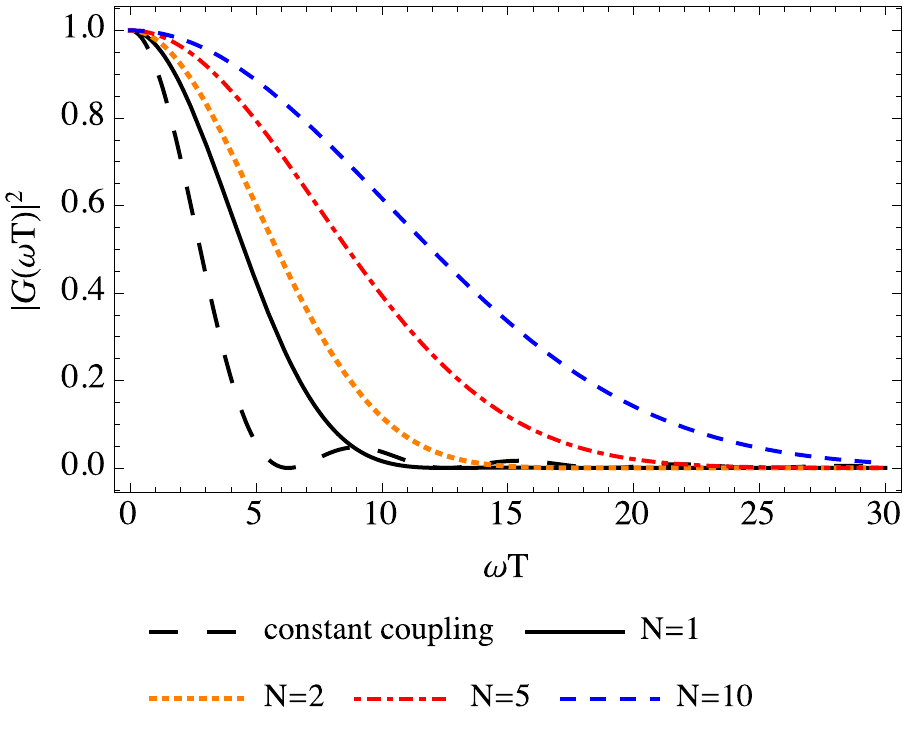}
\caption{\label{fig:decay}(Color online) Squared Fourier transform $\abs{G(\omega T)}^2$ of the coupling functions $g(t)$ displayed in Fig.~\ref{fig:g}, shown as a function of the dimensionless parameter $\omega T$. $\abs{G(\omega T)}^2$ is proportional to the transition probability measuring the state disturbance. The case of constant coupling $g(t)=1/T$ is indicated for comparison. The rapid oscillations of $\abs{G(\omega T)}^2$ are disregarded. (a) Behavior for large values of $\omega T$, the relevant regime for protective measurement. (b) Behavior for small values of $\omega T$, showing the modest increase in width with $N$.}
\end{figure}

Figure~\ref{fig:g} displays the coupling functions determined from the conditions~\eqref{eq:conds} for different values of $N$. Functions with larger $N$ describe a smoother turn-on and turnoff behavior. Figure~\ref{fig:decay}(a) shows the corresponding squared Fourier transforms $\abs{G(\omega T)}^2$ of these coupling functions in the regime $\omega T \gg N$ relevant to protective measurement, with $\abs{G(\omega T)}^2$   representing the dependence of the state disturbance on the choice of $g(t)$. We have neglected the rapid oscillations of $\abs{G(\omega T)}^2$, since they are irrelevant to considerations of state disturbance in protective measurements \footnote{Targeting the zeros of $\abs{G(\omega T)}^2$ to minimize the state disturbance by tuning $T$ would require precise knowledge of $\omega$ and therefore of $\op{H}_S$. But in a protective measurement, $\op{H}_S$ is \emph{a priori} unknown \cite{Aharonov:1993:jm,Dass:1999:az}.}. Small values of $N$ already achieve a strong reduction of the state disturbance.  Figures~\ref{fig:g} and \ref{fig:decay}(a) show that while increasing $N$ entails a higher rate of change of the measurement strength outside the turn-on and turnoff region and a larger peak strength at $t=T/2$, it nevertheless reduces the state disturbance. This indicates that the smoothness of the turn-on and turnoff of the interaction has a decisive influence on the state disturbance.  

Increasing $N$ also makes $g(t)$ narrower (see Fig.~\ref{fig:g}), making its Fourier transform wider and the initial decay of the transition amplitude slower, as seen in Fig.~\ref{fig:decay}(b). However, Fig.~\ref{fig:decay}(a) shows that this increase in width is insignificant in the relevant regime $\omega T \gg N$. Fundamentally, if $N \rightarrow \infty$, $g(t)$ becomes infinitely narrow and the transition amplitude becomes infinitely wide. Thus, one cannot eliminate the state disturbance altogether even in the limit of infinitely many $f_n(t)$. 

We now make precise the connection between smoothness and state disturbance. Mathematically, smoothness is measured by how many times a function is continuously differentiable over a given domain; we call a function that is $k$ times continuously differentiable a $C^k$-smooth function. The $j$th-order derivative of $g(t)=\frac{1}{T} \left( 1 + \sum_{n=1}^N a_n f_n(t)\right)$ [Eq.~\eqref{eq:bvdhkjbvd}] at $t=0$ and $t=T$ is proportional to $\sum_{n=1}^N a_n (2\pi n)^j$ for even $j$ and zero for odd $j$. Since all derivatives of $g(t)$ vanish for $t<0$ and $t>T$, the turn-on and turnoff points introduce a discontinuity in the derivatives. We can make all derivatives up to order $2N-1$ vanish (and thus continuous) at $t=0$ and $t=T$ by requiring that $\sum_{n=1}^N a_n (2\pi n)^{2k}=0$ for $k=1,2,\hdots,N-1$, in addition to the requirement $\sum_{n=1}^N a_n=1$ ensuring continuity of $g(t)$ itself. These, however, are precisely the conditions~\eqref{eq:conds} previously derived from the requirement of eliminating lower-order terms in the Fourier transform. Thus, increasing $N$ makes $g(t)$ arbitrarily smooth, resulting in a polynomial decay of the transition probability to arbitrary order in $1/\omega T$.

\section{\label{sec:minim-state-dist}Minimization of state disturbance using bump coupling functions}

\begin{figure}
\includegraphics[scale=.85]{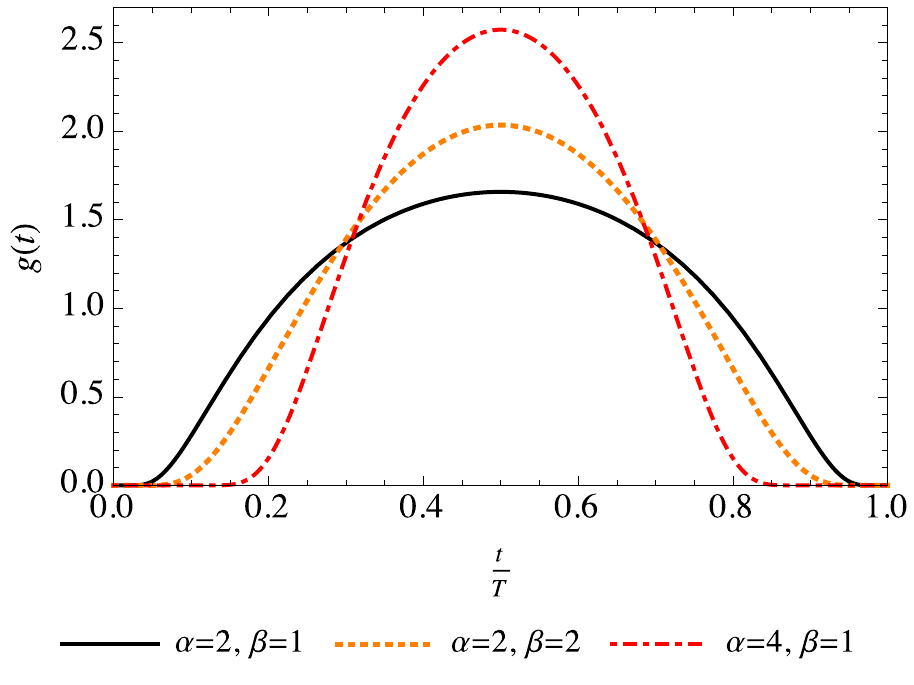}
\caption{\label{fig:gbump}(Color online) Bump coupling functions $g_{\alpha\beta} (t)$ as given by Eq.~\eqref{eq:bumfcts}, shown for different choices of the parameters $\alpha$ and $\beta$. The horizontal axis is in units of the measurement time $T$ and the vertical axis is in units of $1/T$.}
\end{figure} 

The construction of coupling functions from Eq.~\eqref{eq:bvdhkjbvd} progressively increases smoothness and illuminates the relationship between smoothness and state disturbance. However, the decay of the corresponding transition probability with $T$ is only polynomial. This raises the question of whether coupling functions exist that achieve superpolynomial decay. Clearly, this will require functions with compact support $[0,T]$ that are $C^\infty$-smooth, known as bump functions \cite{Lee:2003:oo}.  No such function can have a Fourier transform that follows an exponential decay in $1/\omega T$, since a function whose Fourier transform decays exponentially cannot have compact support. Thus, the state disturbance can at most exhibit subexponential decay. A suitable class of bump functions with support $[0,T]$ is given by 
\begin{equation}\label{eq:bumfcts}
g_{\alpha\beta} (t) = \begin{cases} 
c_{\alpha\beta}^{-1}\exp\left(-\beta \left[1-\left(\frac{2t}{T}-1\right)^2\right]^{1-\alpha}\right), \\ \qquad \qquad \qquad \qquad \qquad \qquad \qquad 0 < t <T,  \\ 0 \qquad \qquad \qquad \qquad \qquad \qquad \quad \,\,\,  \text{otherwise},
\end{cases} 
\end{equation}
where $\alpha \ge 2$ and $\beta\ge 1$ are integers, and $c_{\alpha\beta}$ normalizes the area under $g_{\alpha\beta} (t)$. These functions are $C^\infty$-smooth with vanishing derivatives and essential singularities at $t=0$ and $t=T$. Figure~\ref{fig:gbump} shows $g_{\alpha\beta} (t)$ for several different choices of $\alpha$ and $\beta$.

\begin{figure}
\includegraphics[scale=.9]{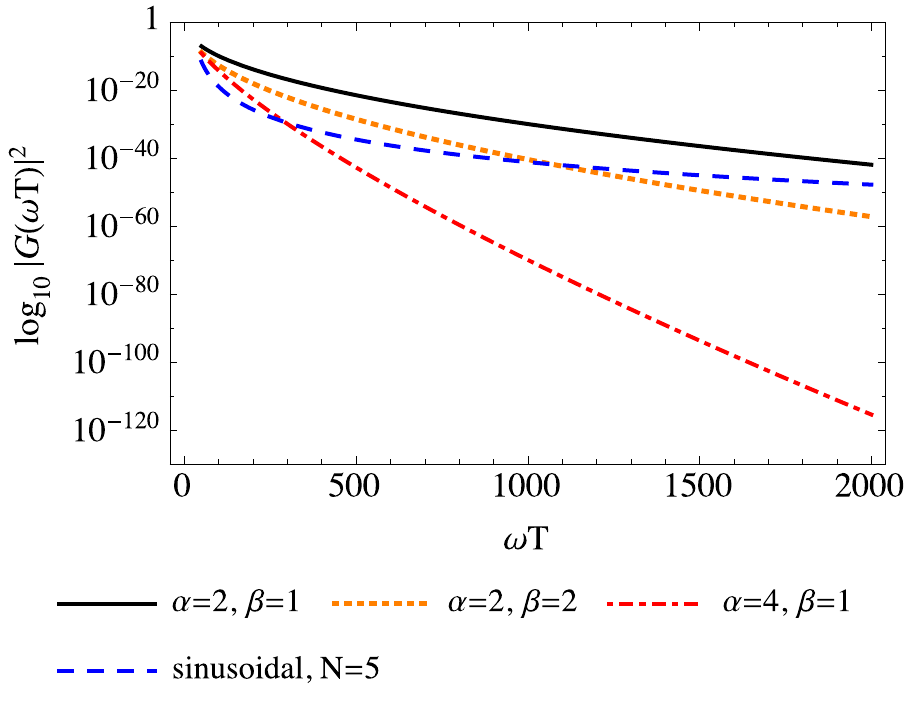}
\caption{\label{fig:bumpsd}(Color online) Squared Fourier transform $\abs{G(\omega T)}^2$ of the bump coupling functions displayed in Fig.~\ref{fig:gbump}, shown as a function of the dimensionless parameter $\omega T$. The result for a coupling function constructed from $N=5$ sinusoidal components [Eq.~\eqref{eq:bvdhkjb678678}] meeting the conditions~\eqref{eq:conds} is shown for comparison. The rapid oscillations of $\abs{G(\omega T)}^2$ are disregarded.}
\end{figure}

For $\alpha=2$ and $\beta=1$, the Fourier transform exhibits subexponential decay proportional to $(\omega T)^{-3/4}\E^{-\sqrt{\omega T}}$ (Fig.~\ref{fig:bumpsd}). Increasing $\alpha$ and $\beta$ enhances the decay (see again Fig.~\ref{fig:bumpsd}), with Fourier transform (to leading order in $1/\omega T$) proportional to $(\omega T)^{-(\alpha+1)/2\alpha} \exp\left[-\gamma_{\alpha\beta} (\omega T)^{(\alpha-1)/\alpha}\right]$, where $\gamma_{\alpha\beta}$ is a constant. By increasing $\alpha$ we can asymptotically approach exponential decay. As seen in Fig.~\ref{fig:gbump}, this will also make $g(t)$ more narrow, rendering the initial decay less rapid, just as for $g(t)$ constructed from an increasing number of sinusoidal components. Figure~\ref{fig:bumpsd} makes clear that since $\omega T \gg 1$, bump functions are superior to coupling functions composed of the sinusoidal components defined in Eq.~\eqref{eq:bvdhkjb678678}.

\section{\label{sec:suff-first-order}Sufficiency of the first-order amplitude}

The higher-order corrections $A_m^{(\ell \ge 2)}(T)$ [Eq.~\eqref{eq:g8fbvsv1}] are of $\ell$th order in the interaction strength, but in general contain terms of first order in $1/T$ \cite{Schlosshauer:2014:pm}. This raises the question of whether the conditions~\eqref{eq:conds}, which eliminate terms up to order $(\omega T)^{-(2N+1)}$ in $A_m^{(1)}(T)$, also eliminate these orders in $A_m^{(\ell)}(T)$ for all $\ell \ge 2$. We find that this is indeed the case. Evaluating $A_m^{(\ell)}(T)$ for $g(t)$ with $N$ nonzero coefficients $a_n$ satisfying the $N$ conditions~\eqref{eq:conds} gives, to leading order in $1/\omega T$,
\begin{align}\label{eq:uuuun22243}
\widetilde{A}_m^{(\ell)}(T)&= \left(-\frac{\I}{\hbar}\right)^\ell \frac{\I O_{mn} }{(\ell-1)!} \left[O_{mm} ^{\ell-1}-O_{nn} ^{\ell-1}\E^{\I \omega T}\right] \left(2\pi\right)^{2N} \notag\\ &\quad \times \left(\sum_{n=1}^N a_n n^{2N}\right) \left(\frac{1}{\omega T}\right)^{2N+1}.
\end{align}
Since this is of the same leading order in $1/\omega T$ as the first-order transition amplitude $A_m^{(1)}(T)$ [see Eq.~\eqref{eq:uuuun2233}], the total transition amplitude $A_m(T) = \sum_{\ell=1}^\infty A_m^{(\ell)}(T)$ is also of the same leading order as $A_m^{(1)}(T)$.

We establish a stronger result still. We calculate the total transition amplitude to leading order in $1/\omega T$ by summing Eq.~\eqref{eq:uuuun22243} over all orders $\ell$. The result is 
\begin{align}\label{eq:uuudfb43}
\widetilde{A}_m(T) &\approx  -\frac{2\I}{\hbar} O_{mn} \E^{\I\omega T/2} 
\sin  \left\{ \frac{\omega T }{2} \left[1 +\chi_{mn}(T) \right]\right\} \left(2\pi\right)^{2N} \notag\\ &\quad \times\left(\sum_{n=1}^N a_n n^{2N}\right) \left(\frac{1}{\omega T}\right)^{2N+1},
\end{align}
where $\chi_{mn}(T) = (\hbar\omega T)^{-1} \left[O_{mm} - O_{nn}\right]$ \footnote{Eq.~\eqref{eq:uuudfb43} omits an overall phase factor, which does not influence the transition probability.}. Comparison with Eq.~\eqref{eq:uuuun2233} shows that the corrections $A^{(\ell \ge 2)}_{m}(T)$ merely introduce a scaling factor $1+\chi_{nm}(T)$ into the argument of the sine function, whose oscillations, however, may be disregarded (see note~\cite{Note2}).  Hence we may replace the sine function by 1, in which case Eqs.~\eqref{eq:uuuun2233} and \eqref{eq:uuudfb43} become identical. Thus, to leading order in $1/\omega T$, the first-order transition probability $\abs{A^{(1)}(T)}^2$ accurately describes the state disturbance. This offers an important calculational advantage and enables the analysis of state disturbance in terms of properties of Fourier-transform pairs.

\section{Discussion}

A particularly intriguing application of protective measurement is the possibility of characterizing the quantum state of a single system from a set of protectively measured expectation values. While this approach is intrinsically limited by its requirement that the system initially be in an eigenstate of its Hamiltonian \cite{Aharonov:1993:qa,Aharonov:1993:jm,Dass:1999:az}, it has the distinct conceptual and practical advantage of not requiring ensembles of identically prepared systems, in contrast with conventional quantum-state tomography based on strong \cite{Vogel:1989:uu,Dunn:1995:oo,Smithey:1993:lm,Breitenbach:az:1997,White:1999:az,James:2001:uu,Haffner:2005:sc,Leibfried:2005:yy,Altepeter:2005:ll,Lvovsky:2009:zz} or weak \cite{Lundeen:2011:ii,Lundeen:2012:rr,Fischbach:2012:za,Bamber:2014:ee,Dressel:2011:au} measurements. Thus, it provides an important alternative and complementary strategy for quantum-state measurement \cite{Aharonov:1993:qa,Aharonov:1993:jm,Aharonov:1996:fp,Dass:1999:az,Vaidman:2009:po,Auletta:2014:yy,Diosi:2014:yy,Aharonov:2014:yy}.

To successfully characterize the initial state of the system with protective measurements, it is crucial that the initial state of the system is minimally disturbed during the series of protective measurements that determine the set of expectation values. We have shown how one can minimize this state disturbance, given a fixed duration $T$ and average strength ($\propto 1/T$) of each protective measurement. Specifically, we have described a systematic procedure for designing the time dependence of the system--apparatus interaction (described by the coupling function) such that the state disturbance decreases polynomially or subexponentially with $T$. The leading order in $1/T$ can be made arbitrarily large for polynomial decay, and one may also come arbitrarily close to exponential-decay behavior by using bump functions. Since strictly exponential decay cannot be attained, bump functions are the optimal choice, as they produce the least possible state disturbance in a protective measurement. 

Previous discussions of protective measurement \cite{Aharonov:1993:qa,Aharonov:1993:jm,Aharonov:1996:fp} have appealed to the condition that the coupling function change slowly during the measurement such that the quantum adiabatic theorem \cite{Born:1928:yf} can be applied. But our results indicate that this condition is both too weak and too strict. It is too weak, because it concerns only the smallness of the first-order derivative of the coupling function, rather than the number of continuous derivatives. It is too strict, because our analysis shows that the state disturbance in a protective measurement is chiefly due to discontinuities in the coupling function and its derivatives during the turn-on and turnoff of the measurement interaction. Once a sufficiently smooth turn-on and turnoff is achieved, the interaction strength may be changed comparably rapidly during the remaining period without creating significant additional state disturbance. Thus, the reduction of the state disturbance through an optimization of the coupling function does not necessitate adjustment of the measurement time or average interaction strength. Furthermore, compared to the condition of smoothness, the weakness of the interaction has a small effect on the state disturbance, which depends only quadratically on the average interaction strength. 

The optimization procedure described here is very general, because it solely modifies the time dependence of the coupling function and is independent of the physical details of the system and the apparatus. In particular, it is independent of the Hamiltonian and the measured observable. This raises the question of whether and how one might further improve the fidelity of the state measurement if the specifics of the physical system and measured observables are taken into account. One approach would be to make use of any available partial knowledge of the Hamiltonian of the system. Such knowledge may be used to additionally reduce the state disturbance, since then the system--apparatus interaction can be designed to target the partially known eigenspaces of the Hamiltonian \footnote{Of course, in the limiting case of a completely known Hamiltonian, a projective measurement in the energy eigenbasis permits determination of the state of the system without any state disturbance, since the system is assumed to be in one of these eigenstates.}. In this case, one may also be able to reduce particular transition amplitudes by minimizing some of the transition matrix elements $O_{mn}=\bra{m} \op{O} \ket{n}$ [see Eqs.~\eqref{eq:uuuun22243} and \eqref{eq:uuudfb43}]. However, this approach can be expected to succeed only for a subset of eigenstates and very few particular choices (if any) of observables $\op{O}$, while state determination requires the protective measurement of multiple complementary (and practically measurable) observables. 

In summary, we have shown how to optimally implement protective measurements and thereby maximize the likelihood of success of protective measurements that seek to determine the quantum state of single systems. Our results dramatically improve the performance of protective measurements and may aid in their future experimental realization.


%

\end{document}